\documentclass
[prl,10pt,letterpaper,twocolumn,bibnotes,notitlepage,final,superscriptaddress,balancelastpage]{revtex4}

\begin{document}

\title{``Local Instruction-Set Model for the Experiment of Pan \emph{et al}'' of 
Aschwanden, Philipp, Hess, and Barraza-Lopez Cannot Beat Quantum Models}
\date{\today}

\author{Wies{\l}aw Laskowski}
\affiliation{Instytut Fizyki Teoretycznej i Astrofizyki, 
Uniwersytet Gda\'nski, 
PL-80-952 Gda\'nsk}

\author{Jian-Wei Pan}
\affiliation{Hefei National Laboratory for Physical Sciences at Microscale and Department
of Modern Physics, University of Science and Technology of China, Hefei, Anhui
230026, China}
\affiliation{Physikalisches Institut, Universit\"{a}t Heidelberg, Philosophenweg 12,
D-69120 Heidelberg, Germany}

\author{Tomasz Paterek}
\affiliation{Instytut Fizyki Teoretycznej i Astrofizyki, 
Uniwersytet Gda\'nski, 
PL-80-952 Gda\'nsk}

\author{Marek \.Zukowski}
\affiliation{Instytut Fizyki Teoretycznej i Astrofizyki, 
Uniwersytet Gda\'nski, 
PL-80-952 Gda\'nsk}

\begin{abstract}
We show the critique of the Pan {\em et al} experiment, given in quant-ph/0503108, is unfounded.

\end{abstract}

\maketitle

The paper \cite{APHB-L} is intended to show, 
that the results of the GHZ-type experiment of ref. \cite{PBDWZ}
cannot prove impossibility of local realism.
The data of the experiment violate Mermin's Bell-type inequality \cite{MERMIN}.
The inequality is in the form $B\leq2$, where $B$ is a linear combination of values of three particles correlation functions, and the bound $2$ holds for the predictions based on locality and realism.
One has $B\leq4$ for any theory - this bound is achievable with {\em pure} Greenberger-Horne-Zeilinger (GHZ) quantum states.
The experimenters found the visibility of the three-particle correlations at the level of $71\%$,
which corresponds to the value of Mermin's expression equal to $2.84$,
which is a drastic violation of the limit for local realistic theories.
In the paper \cite{APHB-L} authors explicitly construct a local realistic model which
saturates the inequality and claim their model better describes experimental results than quantum mechanics. Therefore they declare:
``As a consequence the experimental results can not be used to support [...] quantum nonlocality [...]''\footnote{Both the authors of \cite{APHB-L} and \cite{PBDWZ}
use the term ``quantum nonlocality" as synonymous with violation of local realism. This unfortunate terminology is widespread by now. 
Violations of Bell inequalities falsify local realism, and therefore falsify either realism, or locality, or finally the ``free will" assumption (see, e.g., \cite{GILL}), or a conjunction of all of these assumptions or some of them. Non-locality is not a necessary consequence of violations of Bell inequalities, neither is the mysterious quantum non-locality. Non-locality must follow, if one assumes that realism and ``free will"  hold. However neither quantum mechanics is built basing on the assumption of realism, nor realism is a consequence of quantum formalism.}. 

Let us now present why one cannot agree with the argumentation in \cite{APHB-L}. 
First of all, violation of Mermin's inequality is a sufficient condition for impossibility
of {\em any }local realistic description whatsoever for the observed correlations.
Since the experimental results give the value of the Mermin's expression $B_{exp}= 2.84$, any 
local realistic description of the experiment has been disproved, including the model of \cite{APHB-L}. 
Second, the claim that predictions of the model of \cite{APHB-L} give a
better explanation of experimental data is not due to some failure of quantum mechanics,
but due to a specific rigged choice of quantum model with which the data are compared.
The point is, that in the experiment \cite{PBDWZ} only an approximation to the GHZ state was produced,
and to give a quantum model for the data obtained one must take this into account. The authors of \cite{APHB-L} compare the data and their model 
with the predictions for a perfectly pure GHZ state. Indeed the local realistic model is closer to the data than the perfect GHZ correlations.
However, this does not mean that there does not exist a quantum state that almost perfectly reproduces the data, and therefore describes the results much much better than any local realistic model. 
For example, one can take the mixed state
\begin{equation}
\rho = 0.71 ~ \rho_{GHZ} + 0.29 ~ \frac{1}{8} \openone,
\end{equation}
where $\rho_{GHZ}$ is the projector onto the GHZ state, and $\frac{1}{8} \openone$ is a ``random noise'' admixture, which models the imperfections in the pure GHZ entanglement preparation.
Since for the noise the mean value of any spin or polarization observables is zero, this quantum model
explains the experimental data quite well.
We quote \cite{APHB-L} again:
``Whichever model comes closer to the observed data must be considered to be the better physical model''.
Well, we could not  agree more, but as can be seen from our example, the best physical model to date is quantum mechanics 
(results of which cannot be described within the assumptions of locality and realism). 
This is not only due to the fact that the predictions of this quantum model are 
``optically'' much closer to the experimental results than those for the local-instruction model (we use here the type of intuitive data analysis used in ref. \cite{APHB-L}, see the figure in the reference), 
but also the value of Mermin's parameter, for the state $\rho$, is  $2.84$, 
exactly as the one obtained by Pan {\em et al} 
(something absolutely impossible for any local realistic model).

Let us end with two more remarks.
Of course in the experiment several standard loopholes were present,
but these are common to most of Bell-type experiments, 
and in \cite{APHB-L} they are not discussed, and therefore we neither address them here.
Some specific interpretational problems of the experiment, which seem to be loopholes, but can be shown not to be of any fundamental nature, were discussed in \cite{ZUKOWSKI}.
The reader should also note that in \cite{APHB-L}, 
in contradiction with earlier papers of some of the authors of  \cite{APHB-L}, for example the famous \cite{HESS},
Bell's assumptions of local realism, which lead to his inequalities, are not directly challenged. 
We respond to the new claims. 
The discussion in \cite{HESS} was shown to be 
incorrect earlier in \cite{GILL}.

This note is supported by the KBN grant 1~P03B~04927.
MZ is supported by the Professorial Subsidy of FNP.
WL and TP are supported by stipends out of this Subsidy.

\end{document}